\documentclass{sf2a-conf2018}
\usepackage{graphicx}
\usepackage{hyperref}
\usepackage[]{natbib}  
\usepackage{epstopdf}
\usepackage{amssymb}
\usepackage{amsmath}

\def\BibTeX{{\rm B\kern-.05em{\sc i\kern-.025em b}\kern-.08em
    T\kern-.1667em\lower.7ex\hbox{E}\kern-.125emX}}
\bibpunct{(}{)}{;}{a}{}{,}  

\begin{document}

\TitreGlobal{SF2A 2018}


\title{High energy spectral study of the black hole Cygnus X-1 with INTEGRAL}

\runningtitle{SF2A 2018}

\author{Floriane Cangemi}\address{Laboratoire AIM, UMR 7158, CEA/DSM CNRS Universit\'e Paris Diderot, IRFU/DAp, F-91191 Gif-sur-Yvette, France}

\author{J\'er\^ome Rodriguez$^1$}

\author{Victoria Grinberg}\address{Institut f\"ur Astronomie und Astrophysik, Universit\"at T\"ubingen, Sand 1, 72076 T\"ubingen, Germany} 
\author{Philippe Laurent}\address{Laboratoire APC, UMR 7164, CEA/DSM CNRS Universite\'e Paris Diderot, Paris, France} 
\author{Joern Wilms}\address{Dr. Karl Remeis-Observatory and Erlangen Centre for Astroparticle Physics, Sternwartstr. 7, D-96049 Bamberg, Germany} 
\setcounter{page}{237}


\maketitle


\begin{abstract}
We present the analysis of an extended \textit{INTEGRAL} dataset of the high-mass microquasar Cygnus X-1. We first classify, in a model-independent way, 
all the \textit{INTEGRAL} individual pointings taken between 2003 and 2016 in three basic spectral states. This, in particular, allows us to triple the exposure time 
of the soft state in comparison with previous publication. We then study the spectral properties of 
the 5--400 keV stacked spectra of the soft and hard states and provide the parameters obtained with our modelling. Using a refined alternative method
of extracting the Compton double events of the IBIS telescope, we then extract high-energy ($>$400 keV) spectra in the two states. We do detect an
hard tail in both states. Our refined analysis allows us to obtain a hard state (count) spectrum  at a  flux lower than previously published by our team. Although 
a full estimate of the calibration property of this improved software is still needed, this seems to be more inline with the hard state hard tail seen with other instruments. 
\end{abstract}



\section{Introduction}
Microquasars are variable objects which can transit through several spectral states. The two main ones are the Low Hard State (LHS) and the High Soft State (HSS).  In the LHS, the flux is dominated by emission in the hard X-rays (peaking at $\sim 100$~keV) and by the presence of correlated compact radio jets \citep[e.g][]{corbel_universal_2013-1}. The spectrum is well represented by a power law with a photon index $\Gamma \leqslant 2.0$ and an exponential cutoff around 100 keV. This is usually interpreted as the signature of inverse Compton scattering of soft photons from a (usually undetected) cold disc with hot electrons 
  from a \textquotedblleft corona\textquotedblright (basically described as a hot plasma with an unknown geometry). On the other hand, the HSS is dominated by emission in the soft X-rays, 
  dominated by a black body-like spectrum which peaks around 1 keV,  attributed to an optically thick and geometrically thin accretion disc. No jets are 
  detected in this state but a non-thermal component (with $ \Gamma > 2.5$) can also be present \citep{remillard_x-ray_2006}. 
In addition to these, in both states, a hard powertail is sometimes observed above 400 keV \citep[e.g][]{2000ApJ...543..928M, laurent_polarized_2011, rodriguez_spectral_2015} which could extend to the GeV in some cases \citep[e.g.][]{bodaghee_gamma-ray_2013, loh_high-energy_2016}.
While the twenty past years have permitted to make some advances in the global understanding of these sources' behaviour,  many questions still remain: in particular \textit{how sources transition from a state to another? What causes the transitions?  What are the links between accretion and all type of ejections? 
What is the origin and the nature of the hard powertail at higher energies? How does it rely on the accretion-ejection properties?}
  
The famous and well studied high mass X-ray binary Cygnus X-1 is one of the best candidate to try and answer these questions. 
It is one of the brightest persistent source in the X-rays domain and resolved radio jets have been observed \citep{stirling_relativistic_2001}. The system is close and situated at $d=1.81 \pm 0.01$ kpc \citep{reid_trigonometric_2011} with an inclinaison $i=27.1 \pm 0.8$°  \citep{orosz_mass_2011} and an orbital period of 5.6 days. 
It is known to transit  in  (at least) the two spectral states described above and a high-energy hard tail has been observed with several instruments 
(\textit{CGRO/Comptel}, \cite{2000ApJ...543..928M}; \textit{INTEGRAL/IBIS}  in both spectral states \cite{laurent_polarized_2011} and \cite{rodriguez_spectral_2015}; \textit{INTEGRAL/SPI}, 
\cite{jourdain_integral_2014}).  

In this paper, we gather all the \textit{INTEGRAL} data of Cygnus X-1 in order to study the spectral characteristics of the source in each state. The use of 
the \textit{Compton mode} \citep{forot_compton_2007} available thank to the double layer detection of the telescope \textit{INTEGRAL/IBIS} allows us to study 
the high-energy ($>250$~keV) spectrum and probe the properties of the hard tail. This work follows previous studies by our group: in \citet{laurent_polarized_2011} we reported the detection of a high-energy tail from all \textit{INTEGRAL/IBIS} stacked observations (until 2008), and, later,  \citet{rodriguez_spectral_2015} separate all data until 2012 into states before reporting the detection of a  highly polarized high-energy tail in the LHS only. 
Our motivations here are: (1) add new data to the overall sample and in particular extend the observations in the HSS; the source has indeed  transited in this state in 2011 \citep{2011ATel.3616....1G}; (2) use our refined compton-mode analysis with new settings \citep{laurent_integral/ibis_2017}  that have been tested 
on the Crab and V404 Cygni before and that we apply here  on Cygnus X-1.

\section{Observations and spectral classification}

\begin{figure}[ht!]
 \centering
 \includegraphics[width=1\textwidth,clip]{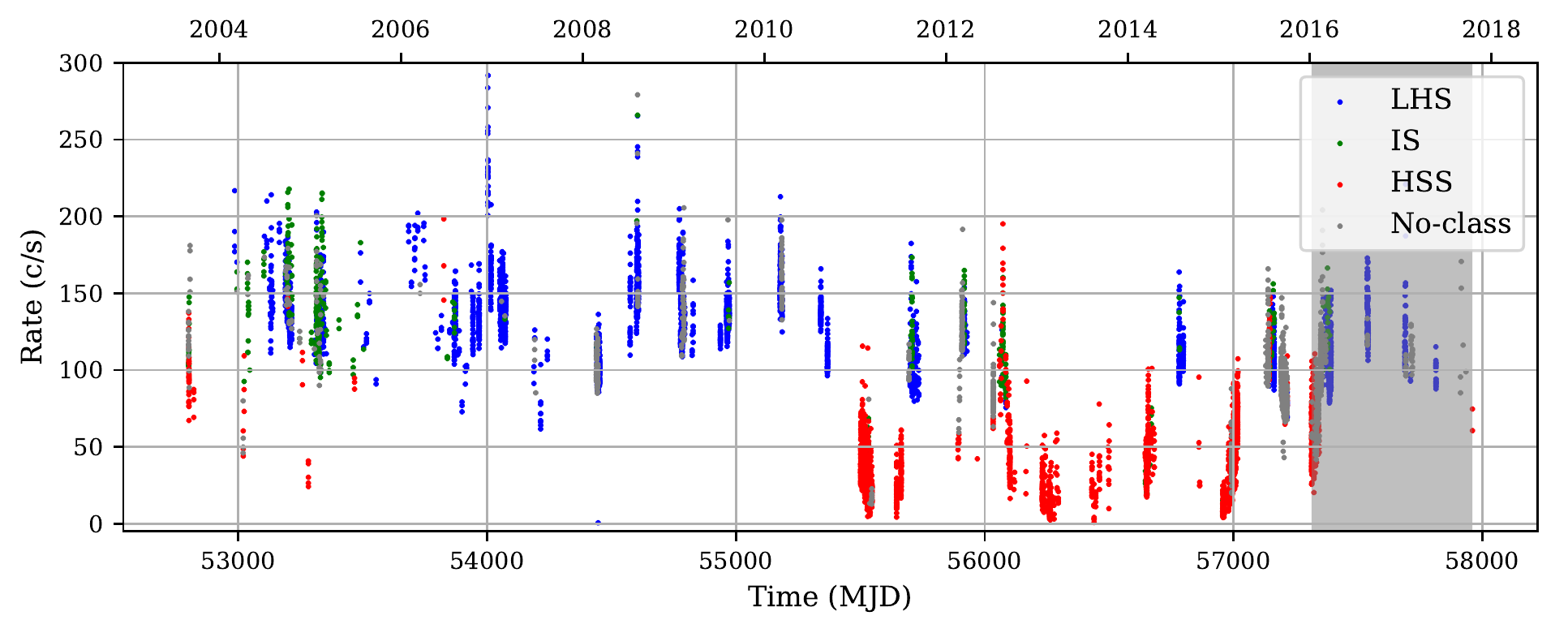}      
  \caption{20--40 keV \textit{INTEGRAL/IBIS}  lightcurve  extracted on a ScWs basis. The blue, green and red dots represent the LHS, IS and HSS ScWs classified according to \cite{grinberg_long_2013}. The grey dots are ScWs that can not be classified with this method. The grey area corresponds to the period after 2016 with dubious calibration. It will be corrected with the realease of the \texttt{OSA version 11} software.}
  \label{cangemi_fig1}
\end{figure}

We used a standard procedure to reduce all the \textit{INTEGRAL/IBIS} data available on Cygnus X-1 from 2003 until today. All details about the extraction 
will be given in Cangemi et al. (2019 in prep). This results in a total of 7122 individual Science Windows\footnote{Individual uninterrupted \textit{INTEGRAL} pointing.}
 (ScWs). Figure \ref{cangemi_fig1} shows the 20--40 keV lightcurve in the  band of these 15 past years, obtained with the \texttt{OSA version 10.2} software. Care should 
 be taken for all fluxes obtained after 2016 (grey area on Figure \ref{cangemi_fig1}), since calibration of the \textit{IBIS} instrument is uncertain.  New version of the
 (\texttt{OSA version 11}) providing good calibration will be released soon (ISDC private communication). All data after 2016 therefore are omitted from 
 the spectral analysis presented here and will be presented in a future publication.

Each ScW has been classified in the same way as we previously applied in the \cite{rodriguez_spectral_2015} using the method described in \cite{grinberg_long_2013}. This method is model-independent and only depends on the countrate and 
hardness ratio of the source as measured by the all sky monitors \textit{RXTE/ASM}, \textit{Swift/BAT} and \textit{MAXI}. We respectively obtain 3156, 544 and 2180 ScWs  in the LHS, Intermediate State (IS) and HSS.  1242 ScWs could not be 
classified with this method and are omitted from the present study.  These may be added after a careful spectral modelling is done. The resulting exposure times 
are 4.89 Ms, 0.90 Ms and 3.31 Ms for the LHS, IS and HSS respectively. 
This extended data sets in particular doubles the exposure times in the LHS and triples it in the HSS compared to our previous work \citep{rodriguez_spectral_2015}. From 2003 until 2011 the source is mainly in the LHS. It then transited in the HSS \citep{2011ATel.3616....1G} and stayed mainly
in this state until late 2015 (Figure \ref{cangemi_fig1}) until it transited back into again the LHS. 

\section{State resolved spectral analysis}

\subsection{Spectrum below 400 keV}
The spectra were obtained from the \textit{INTEGRAL/JEM-X} unit 1 telescope for the energies between 5 and 25 keV and \textit{INTEGRAL/IBIS} for energies 
between 30 and 400 keV. For each states, we stacked all the data before 2016 in order to obtain a state dependent global spectrum. The spectra were
then fitted using \texttt{Xspec 12.9.1} and 2\% of systematics were added.

\begin{figure}[ht!]
 \centering
 \includegraphics[width=0.48\textwidth,clip]{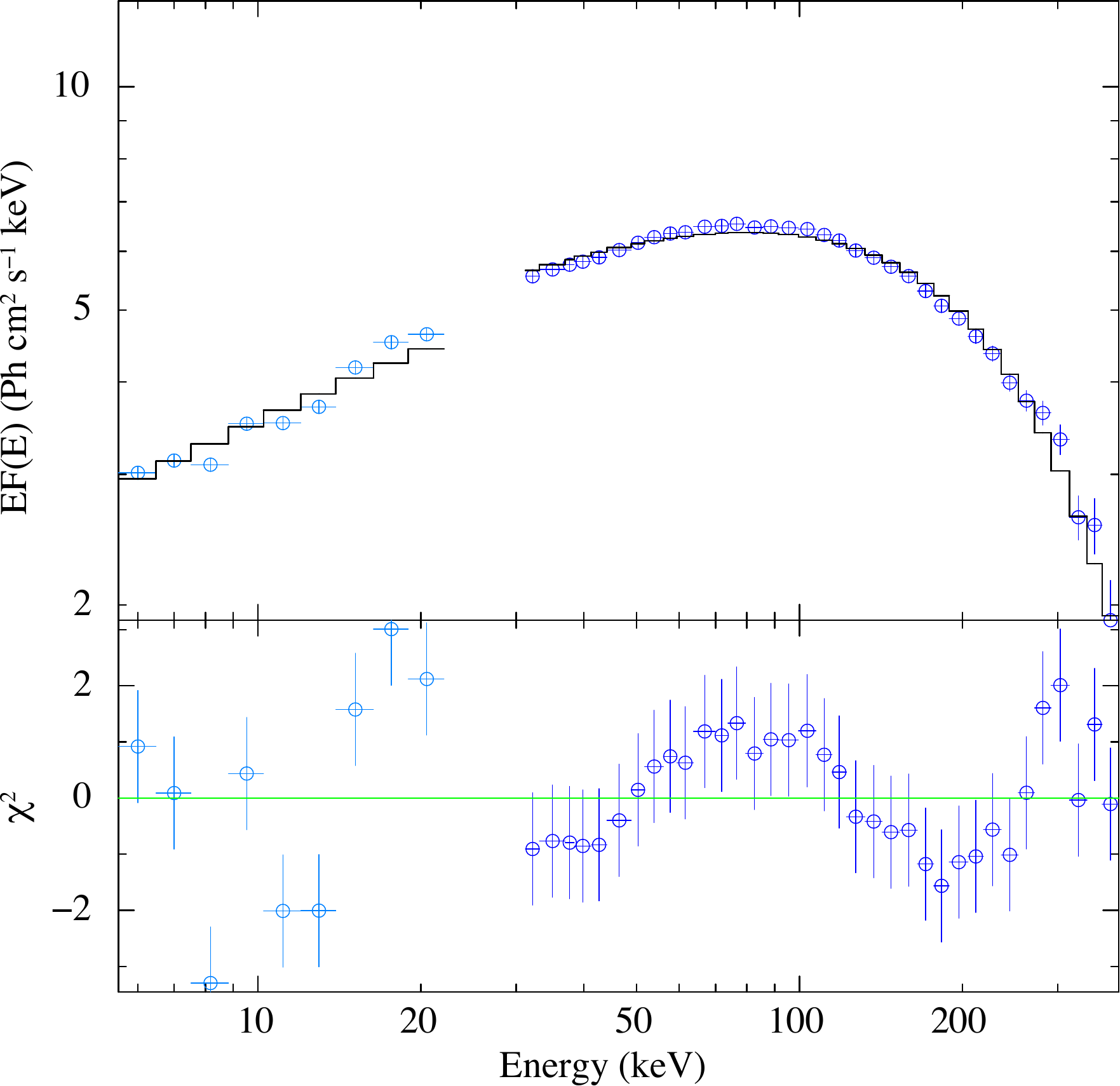}%
 \hspace{0.5cm}
\includegraphics[width=0.48\textwidth,clip]{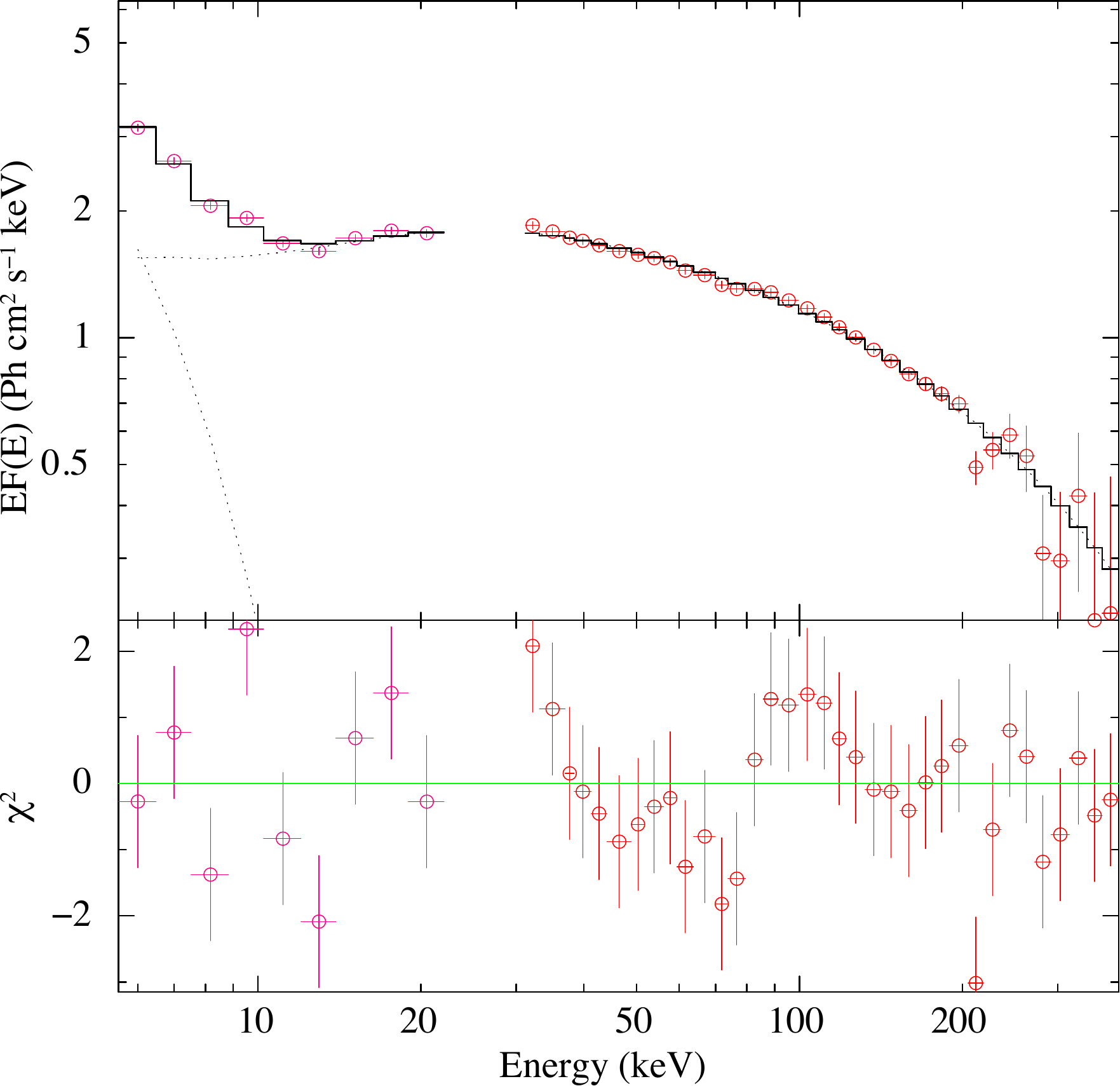}      
  \caption{{\bf Left:} LHS spectrum fitted with \texttt{const * compTT}. The deep blue crosses correspond to data extracted from \textit{INTEGRAL/IBIS} whereas the light blue crosses are data extracted from the \textit{INTEGRAL/JEM-X} telescope. {\bf {Right:}} HSS spectrum fitted with \texttt{const * reflect * (compTT + diskBB)}. The red crosses correspond to data extracted from \textit{INTEGRAL/IBIS} whereas the pink ones are data extracted from the \textit{INTEGRAL/JEM-X} telescope. The bottom plots for each figure show the $\chi^2$ value.}
  \label{cangemi_fig2}
\end{figure}

In the LHS the spectrum was first fitted  with a model of comptonization (i.e \texttt{compTT} \cite{1994ApJ...434..570T}), multiplied by a constant to account for cross calibration and 
differences in the effective exposures between both instruments. The result of the fit is shown in Figure \ref{cangemi_fig2}. We obtained a fit with $\chi^2_{red} = 1.72$ for 39 degrees of freedom (dof). The temperature of the coronal electrons is rather well constrained $kT = 67.3^{+3.3}_{-2.8}$ keV with 
an optical depth $\tau = 0.78 \pm 0.05$. The normalisation constant agree within about 12\%. The addition of a reflection component using \texttt{reflect} \citep{magdziarz_angle-dependent_1995} did not 
improve the fit ($\chi^2_{red}=65.40$ for 38 dof), although residuals seem to be present around $\sim$30 to 50 keV (Figure \ref{cangemi_fig2}).

In the HSS, the residuals to the simple {\tt{compTT}} model show the need for additional components. Both reflection and disc models were added,  i.e  the resulting model is \texttt{constant * (reflect * compTT + diskBB)} in the XSPEC nomenclature. The value of $\cos i$ was fixed using the $ i = 27.1$ \citep{orosz_mass_2011} in the reflection model. The result of the fit is shown in Figure \ref{cangemi_fig2}. We obtained a better fit than for the LHS, 
with $\chi^2_{red} = 1.32$ for 37 dof. We found $kT_{disc} = 1.07^{+0.1}_{-0.09} $ keV for the disc temperature and $kT = 132^{+58}_{-26}$ keV, 
$\tau = 0.14 \pm 0.07$ and $rel_{refl}=0.77^{+0.47}_{-0.28}$ for the corona temperature, the optical depth and the reflection component. The constant for the two 
instruments are, here, identical. 

In order to try to constrain the source's spectral properties in a more physical way, and to further probe the origin of the residuals in the LHS, we fitted the spectra 
with the \texttt{pexrav} and \texttt{pexriv} \citep{magdziarz_angle-dependent_1995} models which are a sum of a cutoff-powerlaw and a reflection 
component in a neutral (\texttt{pexrav}) or ionized (\texttt{pexriv}) medium. Here the reflection is treated in a more self-consistent way, although the input model 
is more phenoomenological (that in the comptonised model used above).  A disc component is still added in the HSS. The results are reported in Table \ref{cangemi_table}. The $rel_{refl}$ value found for the LHS is compatible with very weak or no reflection whereas it is better constrained in the HSS.
 
	\begin{table}
	\begin{center}
	\begin{tabular}{ c c c c c c c c c } 
	\hline
	\hline
     &$\chi^2_{red}$  & dof & $\Gamma$ & $E_{fold}$ (keV) & $rel_{refl}$ & $kT_{disk}$ (keV) &$\xi$ & $const$\\
     \hline
     LHS & $1.14$ &$39$ &  $1.52 \pm 0.02$ & $163.3^{+6.2}_{-5.8}$ & $7.2.10^{-17}$ & - & - & $0.97 \pm 0.03$  \\
     \hline
     HSS &$1.45$ & 35 & $2.00 \pm 0.05$ & $222^{+30}_{-32}$ & $0.84^{+0.47}_{-0.30}$ &  $1.1 \pm 0.1$ & $(5.54 \pm 2.9).10^{-7}$ & $0.99^{+0.06}_{-0.03}$ \\
     \hline
     \hline
 
	\end{tabular}
	\caption{\label{cangemi_table}Table of fit results using the \texttt{constant * pexrav} and \texttt{constant * (pexriv + diskBB)} models for the LHS and the HSS respectively.}
	\end{center}
	\end{table}

\subsection{Spectrum above 400 keV}
\begin{figure}[ht!]
 \centering
 \includegraphics[width=0.48\textwidth,clip]{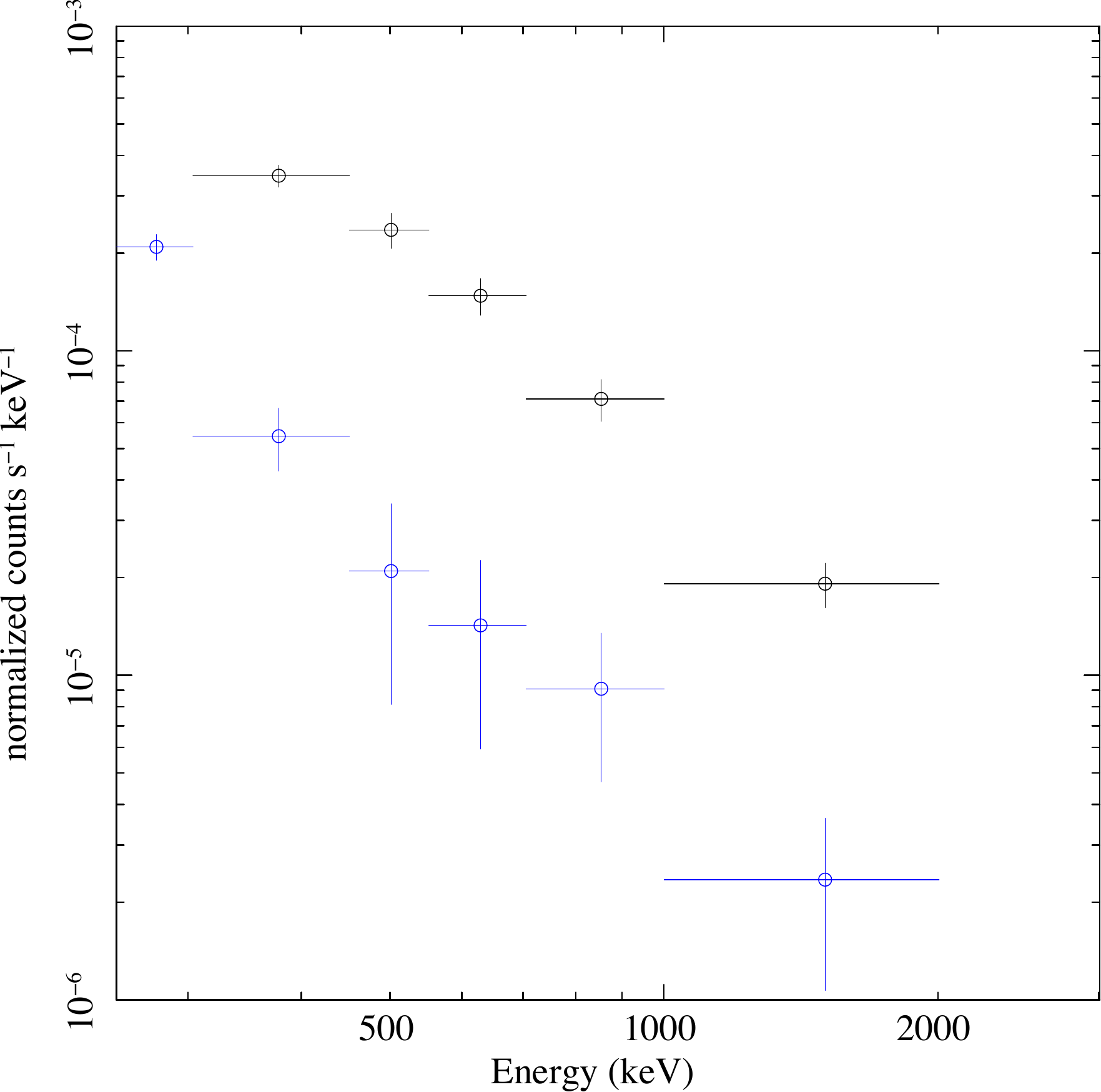}%
 \hspace{0.5cm}
\includegraphics[width=0.48\textwidth,clip]{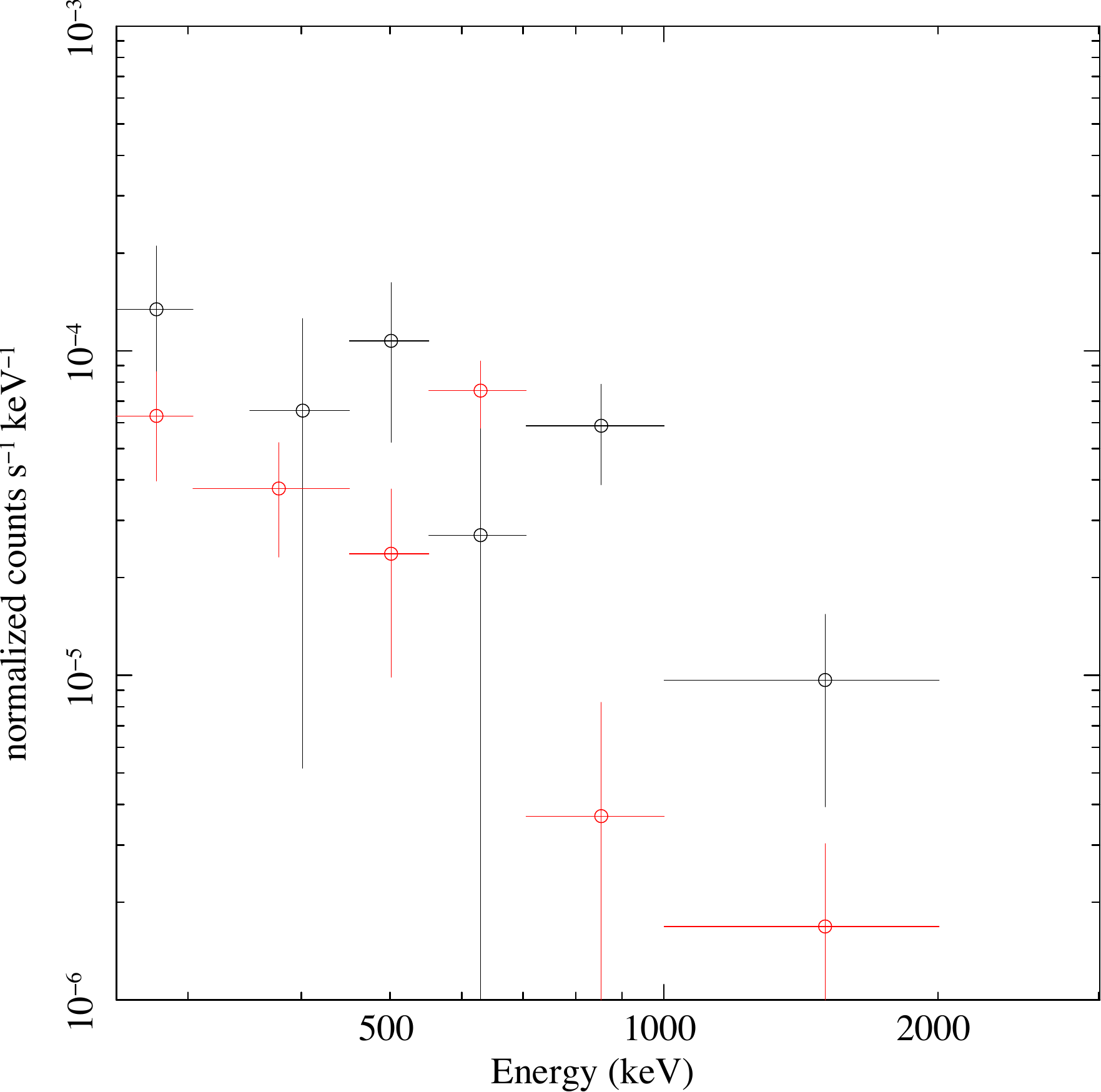}      
  \caption{{\bf Left:} LHS spectra extracted from the \textit{Compton mode}. The blue crosses correspond to the new spectrum obtained with the new method which take into account and substract the \textit{spurious} events whereas the black crosses correspond to the spectrum obtained by \cite{rodriguez_spectral_2015}. {\bf Right}: HSS spectra extracted from the  \textit{Compton mode}. As for the LHS, the red crosses correspond to the new spectrum obtained with the new method which take into account and subtract  the \textit{spurious} events whereas the black crosses correspond to the spectrum obtained by \cite{rodriguez_spectral_2015}.}
  \label{cangemi_fig3}
\end{figure}

The high-energy spectrum is obtained using the \textit{Compton mode} available with the two detector layers of the \textit{INTEGRAL/IBIS} telescope (\cite{forot_compton_2007}, \cite{laurent_polarized_2011}). Some high-energy photons can be detected as double events as they are Compton scattered from \textit{ISGRI} (the upper detector) to \textit{PICsIT} (the lower one)\footnote{For more details  see \cite{forot_compton_2007})}. This method has already been used by \cite{laurent_polarized_2011} and \cite{rodriguez_spectral_2015} to measure the high energy spectrum of Cygnus X-1. At the time, however, some 
discrepancies between the flux measured with IBIS and the fluxes measured with CGRO/Comptel \citep[albeit at totally different epochs of observations][see Figure 5 of \cite{rodriguez_spectral_2015}]{2000ApJ...543..928M}, or even \textit{INTEGRAL/SPI }\citep[although the dataset used was not exactly the same][]{jourdain_integral_2014}. The extreme brightness of V404 Cygni during its 2015 outburst \citep[e.g.][]{rodriguez_correlated_2015-1} lead us to revised and modified 
 the non-standard procedure of extraction of Compton-mode photons \citep{laurent_integral/ibis_2017}. In particular, this modification basically consists of  subtracting all \textit{spurious} events, i.e unrelated events that are detected within the same temporal 
windows, and thus counted as double (this resembles somehow the pile-up effect seen in CCD X-ray detectors for bright sources). The example of V404 Cygni \citep{2016int..workE..22L}, thereafter applied to the Crab Nebula \citep{2016int..workE..38G}, showed that they could contribute somewhat should be corrected. The result of this refined analysis of the Cygnus X-1 data are shown in Figure \ref{cangemi_fig3}. The black spectra are count 
spectra obtained by \cite{rodriguez_spectral_2015}, and the coloured one are the one obtained with the new method. In the LHS a significant diminution 
of the count rate is obvious and shows that the hard tail flux might be slightly lower than the one previously published, although a full characterisation 
of the new instrumental response is needed before we can obtain the definitive spectral parameters. 
More importantly a hard tail is detected, for the first time with \textit{INTEGRAL}, in the HSS up to at least 500 keV.

\section{Discussion and conclusion}
We extracted all \textit{INTEGRAL/JEM-X} and \textit{INTEGRAL/IBIS} Cygnus X-1 data available since the \textit{INTEGRAL} lauch. The model independent state classification based on \cite{grinberg_long_2013} enables us to stack data for each state and obtain averaged spectra over long exposure times. 
We first used a model of thermal Comptonization to fit the LHS spectra, and we added a reflection and disc components 
to model the HSS ones. In the LHS, the high-energy cutoff is well defined at an energy slightly higher than the value obtained by \cite{rodriguez_spectral_2015} with an  opitcal depth slighty lower. This could either be due to the degeneracy of these two parameters or simply indicate small intrinsic variability of the coronal
parameters.  In the HSS we do detect the presence of a $\sim$1 keV accretion disc in addition to a hard tail and a reflection component. When fitting with thermal 
Comptonisation, the temperature of the corona is not well constrained but seems a bit higher than in  \cite{rodriguez_spectral_2015} with a higher optical depth. The reflection properties are essentially similar. All differences could come from intrinsic variations between the two dataset used.
Then we fitted with \texttt{pexrav} and \texttt{pexriv}, phenomenological but auto-consistent models which combine reflection and high-energy cutoff. 
We found no reflection in the LHS but we do measure reflection in the HSS. 
All these results are easily understandable in the context of a truncated disc in LHS that would shrink towards the black hole as the source transits in the soft state. However, these results are in contradiction with other studies in which reflection is detected in the LHS \cite[e.g][]{wilms_long_2006}, and this will be discussed in a future paper.

The spectrum of a population of thermal photons undergoing  thermal Comptonisation in a plasma of temperature $T$ with an optical depth $\tau$,  assuming 
a spherical geometry,  is a powerlaw with flux $F_\nu \propto \nu^{-\alpha}$ \citep{1979Natur.279..506S} where:
\begin{equation*}
\alpha = - \frac{3}{2} + \left[\frac{9}{4} -\frac{\pi^2}{3} + ÷\frac{m_ec^2}{kT \left( \tau+\frac{2}{3}\right)^2} \right]^{\frac{1}{2}}
\end{equation*}
and the photon index $\Gamma = 1 + \alpha$. From the parameters obtained with our {\tt{compTT}} model we deduce found a photon index 
$\Gamma = 1.1 \pm 0.1$ for the LHS and $\Gamma = 1.7^{+1.2}_{-1.0}$ for the HSS. The photon indices are thus lower than the ones obtained 
with \texttt{pexrav}, especially in the LHS. These discrepancies could simply indicate that either the Comptonisation is not purely thermal, and that the corona is 
\textquotedblleft hybrid\textquotedblright or that another component dilute a purely thermal Comptonisation (such as the high-energy powerlaw tail).  In the HSS the uncertainties on the coronal
parameters  do not allow us to draw conclusions since the parameters are compatible within the errors. 

We then looked at higher energy thanks to the \textit{Compton mode}. By extracting the data with a refined software, we obtained a significant diminution 
of the source count flux in the LHS, but still detect a significant hard tail above 400 keV in the data. We do detect a hard in the HSS at least up to 500 keV. 
These are of course preliminary results and an updated response of the \textit{Compton mode} is currently being produced in order to allow us to estimate the 
final spectral parameters of this tail and thus precisely characterise it. 
This aspect is of prime importance and is currently our main priority as the origin and the nature of the hard energy tail is still in debate. Some models suggest that it might the signautre of hybrid thermal/non thermal corona \citep[e.g][]{mcconnell_soft_2002, romero_coronal_2014}. Others invoke synchrotron radiation from the jets as proposed by \cite{laurent_polarized_2011} and \cite{rodriguez_spectral_2015}. In this case, however, the origin of the emission in the HSS is 
necessarily different since no obvious jet emission is detected in this state. A way to discriminate will be the measure of the polarisation properties (or absence thereof) of this hard tail. The large degree of polarisation in the LHS stimulated the jet interpretation. The upper limit obtained in the HSS was too high for 
meaningful constraints to be obtained. Polarisation studies are underway and should bring new results in the future.

\begin{acknowledgements}
Floriane Cangemi, J\'erome R\^odriguez and Philippe Laurent acknowledge financial support from the French Space Agency (CNES).
Victoria Grinberg is supported through the Margarete von Wrangell fellowship by the ESF and the Ministry of Science, Research and the Arts Baden-W\"urttemberg. 
\end{acknowledgements}

\bibliographystyle{aa}  
\bibliography{cangemi} 

%
\end{document}